\documentstyle[aps,eqsecnum]{revtex}

\newcommand{\be}{\begin{equation}}
\newcommand{\ee}{\end{equation}}
\newcommand{\bea}{\begin{eqnarray}}
\newcommand{\eea}{\end{eqnarray}}

\begin{document}
\title{The point-splitting regularization of $(2+1)$d parity breaking models}
\author{D.\ G.\ Barci$^{a,b}$, J.\ F.\ Medeiros Neto$^{c}$, L.\ E.\ Oxman$^{b}$ and
S.\ P.\ Sorella$^{b}$}
\date{November 17, 2000}
\address{$a)$ Department of Physics, University of Illinois at Urbana-Champaign\\
1110, W. Green St., Urbana, IL 61801-3080, USA\\
$b)$Instituto de F\'{\i}sica, Departamento de F\'\i sica Te\' orica\\ 
Universidade do Estado do Rio de Janeiro\thanks{Permanent address \newline e-mail: barci@highgate.physics.uiuc.edu}\\
Rua S{\~a}o Francisco Xavier, 524, 20550-013, Rio de Janeiro, Brazil\\
UERJ/DFT-02/2000\\
$c)$Universidade Estadual de Santa Cruz (UESC), Departamento de F\'{\i}sica\\
Rodovia Ilh\' eus-Itabuna, Km 15 CEP 45650-000, Ilh\' eus, BA, Brazil}
\maketitle

\begin{abstract}
The coefficient of the Chern-Simons term in the effective action for massive
Dirac fermions in three dimensions is computed by using the point-splitting
regularization method. We show that in this framework no ambiguities arise.
This is related to the fact that the point-splitting regularization does not
introduce additional parity breaking effects, implementing one possible
physical criterion in order to uniquely characterize the system.
\end{abstract}
\noindent {\bf Keywords:} Chern-Simons, Point-Splitting, Parity Anomaly

\noindent {\bf Pacs:} 11.10.-z, 11.10.Kk, 11.10.Gh, 11.15.-q

\noindent {\bf Preprint:}  hep-th/0011154
 
\draft


\section{Introduction}


It has long been known that three-dimensional gauge theories with parity or
time reversal symmetry breaking display a set of far-reaching and
interesting properties as, for instance, the celebrated mechanism for
generating massive excitations without gauge symmetry breaking . Also, the
behavior of $(2+1)$ dimensional Dirac fermions in an electromagnetic
background was extensively studied in the context of $QED_3$ as a prototype
to understand $QED_4$ at finite temperature \cite{jackiw}.

In particular, it is well established that the dynamics of $(2+1)$d massive
Dirac fields induces a topological Chern-Simons term whose coefficient is
not renormalized by higher order Feynman diagrams \cite{oneloop,SSW}. It is
worth to point out that the one-loop effective action $S_{eff}(s)$
corresponding to the three-dimensional fermionic determinant has been worked
out by using several regularization methods \cite
{jackiw,Niemi-Semenoff,Reldich,IM,gam,GRS,mat,rothe,chai,cl,hd}. The result
however shows that a certain degree of ambiguity is present, due to the
regularization dependence of the induced Chern-Simons coefficient $\sigma $,
defined by the expressions 
\begin{eqnarray}
S_{eff}[s] &=&\sigma S_{CS}[s]+~{\rm higher~order~terms\,,}  \nonumber \\
S_{CS}[s] &=&\frac 12\int d^3xs_\mu \epsilon ^{\mu \nu \rho }\partial _\nu
s_{\rho \,},
\end{eqnarray}
where $s_\mu $ is an external gauge field. Nonetheless, this coefficient
plays a central role in the bosonization of $(2+1)$d fermionic systems and
in the related applications to condensed matter, so that one is faced with
the problem of its physical determination. We remind that the bosonization
of the $(2+1)$d fermionic action $K_F[\psi ]$ for {\it free} massive
fermions 
\begin{equation}
K_F[\psi ]=\int d^3x\,\bar{\psi}\left( i\partial \!\!\!/+m\right) \psi \;,
\label{faction}
\end{equation}
is implemented in terms of a bosonizing gauge field $A_\mu $, while the
current correlation functions are reproduced by the exact bosonization rule $%
j^\mu =\bar{\psi}\gamma ^\mu \psi \leftrightarrow \epsilon ^{\mu \nu \rho
}\partial _\nu A_\rho $ \cite
{Marino,FF,Fidel,BFO,c1,banerjee,determinant,BOS,Linhares}. Although a
closed form for the bosonized action $K_B[A]$ is not available, it turns out
that $K_B[A]$ is a gauge invariant functional whose leading term is the
Chern-Simons action with coefficient given by $\sigma ^{-1}$. This is the
dominant term when a large mass expansion is considered.

Recently, we have been able to show \cite{BOS} that the above mentioned
current bosonization rule is not only exact but is also universal. This
means that when bosonizing a fermionic system containing current
interactions $I[j^\mu ] $, the correlation functions can be obtained from
the mapping 
\begin{equation}
K_F[\psi ]+I[j^\mu ]+\int d^3xs_\mu j^\mu \leftrightarrow K_B[A]+I[\epsilon
^{\mu \nu \rho }\partial _\nu A_\rho ]+\int d^3xs_\mu \epsilon ^{\mu \nu
\rho }\partial _\nu A_\rho \;.  \label{urules}
\end{equation}
This relationship has to be understood as an equivalence among the partition
functions defined by the left (fermionic) and right (bosonic) sides, where $%
K_B[A]$ is the functional corresponding to the bosonization of the {\it free}
fermionic action $K_F[\psi ]$.

In ref.\onlinecite{universal}, using the mapping (\ref{urules}), we studied
transport properties in two-dimensional systems presenting a charge gap and
displaying parity breaking properties. There, we related the universal rules
(\ref{urules}) to the universal character of the transverse conductance $I_t$
between two ``perfect Hall regions'', that is, regions where the parity
breaking parameter goes to infinity. We remind that in the relativistic
case, the parity breaking parameter is the fermion mass itself. These
perfect Hall regions were supposed to be adiabatically connected to regions
containing arbitrary current interactions $I[j^\mu ]$. In particular, we
have shown that the transverse current between two ``perfect Hall regions''
is given in terms of the electric potential difference $\Delta V$ between
them, according to 
\begin{equation}
I_t=\sigma \;\Delta V\;,  \label{conduc}
\end{equation}
where $\sigma $ is the induced Chern-Simons coefficient coming from the
fermionic determinant. Notice that this result does not depend neither on
the particular geometry of these regions, nor on the current interactions
localized outside them. These considerations apply also to the
nonrelativistic case, where the parity breaking parameter is given by the
external magnetic field \cite{universal}. For example, when the first Landau
band is completely filled, the induced Chern-Simons coefficient is
unambiguously given by $\sigma =1/(2\pi )$, which implies that the value, in
usual units, of the universal Hall conductance is $e^2/h$. We underline that
the topological information encoded in the mapping (\ref{urules}) \cite{BOS}
, together with the particular parity breaking properties of the system, are
all we need to derive the universal behavior of the transverse conductance.
To some extent, the bosonization technique enhances the topological
properties of the fermionic ground state.

Coming back to the relativistic case, as the induced Chern-Simons
coefficient is related to a physical observable (the universal Hall
conductance), further criteria have to be imposed in order to determine the 
{\it a priori} ambiguous value this coefficient can display. Indeed, from a
purely theoretical point of view there is no way to decide which is the
result, as there is no compelling theoretical reason to disregard a given
regularization scheme on the basis of some serious inconsistency.

Therefore, the induced Chern-Simons coefficient has to be determined by
additional physical requirements, to be specified according to the framework
in which the relativistic fermionic action is being considered.

The action (\ref{faction}) is in fact extensively used in the context of the
effective models describing the so called quantum critical transitions \cite
{duality} and for the nodal quasiparticles \cite{fisher}.

In the first example, a fermionic field with Thirring-like interactions has
been considered as a quantum critical model describing the topological
transitions between plateaus in the Integer Quantum Hall Effect (IQHE) %
\onlinecite{duality}. These transitions are characterized by a transverse
conductivity $\sigma_{xy}=(1/2n) (e^2/h)$ ($n$ integer), and a longitudinal
conductivity $\sigma_{xx}$ which is {\it finite}, due to quantum
fluctuations on all length scales. Indeed, any fermionic model displaying
this behavior is called quantum critical; the value of the transverse
conductivity defines the phase transition point.

The second example corresponds to nodal liquid models for high $T_c$
superconductors \cite{fisher}. The excitations here are described by a
couple of $(2+1)$d Dirac fermions. Depending on the order parameter used to
describe the $d-{\rm wave}$ superconductor ($d_{x^2-y^2}$ or $d_{x^2-y^2}+i
d_{xy}$), the system does (or does not) break parity and time reversal
symmetry.

In both examples, a central property characterizing the physical system is
the amount of parity symmetry breaking, which plays a fundamental role in
order to construct the corresponding effective lagrangians and which should
be considered as establishing a possible set up for the determination of the
model. In other words, we should adopt here the criterion of not introducing
additional parity breaking, whenever the theoretical possibility of
ambiguous results shows up at the quantum level. In fact, if this is to be a
possible physical criterion for the determination of the system, then, any
regularization scheme having no additional sources of parity breaking should
lead to the same result.

The aim of this paper is twofold. First, we address the issue of the
determination of the coefficient $\sigma$ within the framework of the
point-splitting regularization, which belongs to the above mentioned class
of regularization schemes. We will follow the method presented in ref. %
\onlinecite{becchi-velo}, where the point-splitting was successfully applied
to study anomalies in non-abelian chiral gauge theories. Among the schemes
which do not introduce additional parity breaking, the point-splitting turns
out to be particularly adapted to the present case, as it combines at any
stage many desirable properties. It can be implemented at the lagrangian
level, and preserves translation and Lorentz symmetry, as well as invariance
under {\it small} gauge transformations. We will see that the obtained Hall
conductance equals half the perfect value ($\frac{1}{2} \frac{m}{|m|} e^2/h$%
), namely $\sigma=\frac{1}{4\pi} \frac{m}{|m|}$, as it is generally assumed
in the framework of condensed matter systems \cite{fradkinbook}.

Second, we have tried to collect a large amount of information about the
coefficient $\sigma $, comparing the results obtained by different
regularizations. This point should be of some usefulness in order to have a
general view of the situation concerning $\sigma $, helping in clarifying
the possible physical criterion to be adopted for its determination.
Although in the following we shall restrict ourselves to the zero
temperature case, we shall also mention briefly the important progress which
has been recently achieved for nonvanishing temperature \cite
{fidfosco,dunne,deser}.

The paper is organized as follows. In section \S \ref{S:zero-modes} we
report on different regularization methods. In section \S \ref{pdet} we
analyze the relationship between the coefficient $\sigma$ and parity
properties. Section \S \ref{S:point-splitting} is devoted to the evaluation
of the induced Chern-Simons coefficient by using the point-splitting.
Finally, in section \S \ref{S:conclusions}, we present the conclusions.


\section{Regularization ambiguities}

\label{S:zero-modes} 

The simplest example for a $(2+1)$d model with parity breaking properties is
the massive Dirac fermion model. In this case, the parity breaking parameter
is the fermion mass; indeed, under a parity transformation ${\cal P}$, the
lagrangian density transforms according to 
\begin{equation}
\bar\psi \left( iD \!\!\!/+m \right) \psi\stackrel{\cal P}{\rightarrow} \bar%
\psi \left(iD \!\!\!/-m \right) \psi.  \label{parity}
\end{equation}
As a consequence, the fermionic effective action, which gives the system's
response to the external electromagnetic field $A_\mu$, is expected to
contain a parity breaking term, that is, an induced Chern-Simons term.
Moreover, the induced Chern-Simons coefficient should be naively expected to
be related to the sign of the fermion mass, since the parity breaking is not
related with the absolute value of $m$ but with its sign (see eq.\ (\ref
{parity})). However, the presence of superficially linear ultraviolet
divergences, when computing the one-loop effective action, require the
introduction of some regularization scheme. Upon a closer look, the induced
Chern-Simons coefficient turns out to be finite but ambiguous, depending on
the particular way we regulate the divergences. Any induced Chern-Simons
term in the effective action whose coefficient is not related with the sign
of the mass is called an {\em anomalous term}, in the sense that it
represents an additional parity breaking which is not initially present in
the classical fermionic action.

Another symmetry, that could be expected to be present in the fermionic
effective action, is the invariance under large gauge transformations, which
arise when the euclidean time coordinate is compactified to a circle in
order to deal with finite temperature. This symmetry would follow from the
path integral definition of the effective action, where all fermion field
configurations are considered, including those corresponding to large gauge
transformations. In this context, a series of recent articles have shown
that anomalous terms, together with non-extensive parity breaking terms, are
required in order to preserve large gauge invariance of the finite
temperature effective action \cite{fidfosco,dunne,deser}.

Let us proceed thus by reviewing the results obtained by different
regularization schemes.


\subsection*{Regularization schemes}


\begin{description}
\item  {a)} {\em Dimensional regularization}

The dimensional regularization has been employed by the authors of ref. %
\onlinecite{IM}. The result is given by

\begin{equation}
\sigma =\frac 1{4\pi }\frac m{|m|}  \label{sigma1}
\end{equation}

We observe that the same result has been obtained by the differential
regularization \cite{chai}.

\item  {b)} {\em $\zeta $-function regularization}

The $\zeta $-function regularization is based on the calculation of the
fermionic current, by means of a regularized Dirac Green function. To this
aim, the regularized current is written as 
\begin{equation}
J_\mu ^{{\rm reg}}=\left. -\frac{d~}{d\lambda }\left\{ \lambda {\rm Tr}%
\left[ \gamma _\mu \left( \partial \!\!\!/+s\!\!\!/+m\right) ^{-\lambda
-1}\right] \right\} \right| _{\lambda =0}
\end{equation}
where $\lambda $ is a complex variable to be analytically continued to $%
\lambda =0$, where the Green function is not well defined. Due to subtleties
in the analytic continuation, the result contains extra parity breaking
anomalous terms which are not associated with the sign of the original
fermion mass. In this case the result is \cite{gam,GRS} 
\begin{equation}
\sigma =\frac 1{4\pi }\left( \frac m{|m|}\pm 1\right)  \label{sigma2}
\end{equation}
These two possibilities have been related to a determination of the system
compatible with invariance under large gauge transformations, when the
euclidean time coordinate is compactified to live on a circle $S^1$.

\item  {c)} {\em Pauli-Villars regularization}

The induced Chern-Simons coefficient evaluated by means of the Pauli-Villars
regularization was presented in several places \cite{Reldich}. The result
obtained is 
\begin{equation}
\sigma =\frac 1{4\pi }(\frac m{|m|}+q),  \label{q}
\end{equation}
where $q$ is an integer. It turns out that in the abelian theory it is
possible to choose $q=0$, to all orders in a perturbative expansion,
provided we take an appropriate coupling constant renormalization\cite
{baboukhadia}.

It is worth spending here some words about the regularization ambiguity
present in the expression for $\sigma $. In fact, the Pauli-Villars
regularization is expected to be related with a higher order derivative
regularization, where each regulating mass parameter is associated to a new
Dirac factor that changes the fermion propagator, so as to regularize the
ultraviolet behavior, namely 
\begin{equation}
(i\partial \!\!\!/+m)^{-1}\rightarrow \left[ (i\partial
\!\!\!/+m)\prod_i(i\partial \!\!\!/+\Lambda _i)/\Lambda _i\right] ^{-1}.
\end{equation}
In general, this procedure changes the parity properties of the starting
action. Each regulating mass $\Lambda _i$ should contribute an additional
term $q_i=\Lambda _i/|\Lambda _i|$ to the induced coefficient. The change in
the parity properties of the initial theory is measured by $q=\sum_iq_i$
(cf. eq.(\ref{q})). Then, if we are interested in working with an effective
action having the same parity properties of the classical fermionic field
theory, we should consider an equal number of positive and negative
regulating masses. This would correspond to set $q=0$ in eq.(\ref{q}). Note
also that the Dirac factor associated to a pair of regulating masses $%
\Lambda $ and $-\Lambda $, 
\begin{equation}
(i\partial \!\!\!/+\Lambda )(i\partial \!\!\!/-\Lambda ),
\end{equation}
(which corresponds to $q=0$) is invariant under a parity transformation, 
{\it i.e.}, it does not introduce additional parity breaking. In this case,
if the external field coupling is maintained to be $A_\mu \bar{\psi}%
(x)\gamma ^\mu \psi (x)$, a systematic iterative procedure to recover gauge
invariance order by order in perturbation theory must be considered \cite{hd}
.

\item  {d)} {\em Lattice regularization}

When the ultraviolet divergences are regulated by defining the theory on a
lattice, the induced coefficient is \cite{cl} 
\begin{equation}
\sigma =\frac 1{4\pi }\frac m{|m|}n
\end{equation}
where $n$ is an arbitrary integer, identified with a topological number \cite
{cl}. It turns out that $n$ is the winding number that appears when the
fermion propagator (in momentum representation) is viewed as a mapping from
a 3-dimensional torus onto the space of $SU(2)$ matrices.

Notice that this result coincides with the Pauli-Villars regularization,
although the interpretation for the integer $n$ is very different. While in
the later case it is easier to relate the ambiguous coefficient with
additional parity breaking, in the former one, the ambiguity is related to
the different possible inequivalent formulations for fermions on the lattice.
\end{description}


\section{Physical determination of the system}

\label{pdet}

In the introduction, we have stressed that the universal transverse
conductance is identified with the induced Chern-Simons coefficient $\sigma$
in the fermionic effective action (cf. eq.\ref{conduc}). However, as
discussed in the previous section, this coefficient is affected by
ambiguities which have to be fixed by a suitable physical criterion. To this
aim, we discuss in this section the relationship between $\sigma$ and
parity. Before going any further, let us underline a key property concerning
the ambiguity, namely, all the results have the following form 
\begin{equation}
\sigma = \frac{1}{4\pi} \times {\rm integer}  \label{integ}
\end{equation}
whatever the particular regularization scheme is, suggesting thus a
topological origin for $\sigma$. In fact, as shown in \cite{mat}, the Ward
identity for small gauge invariance allows to relate $\sigma$ to a
topological invariant which has the form of a Wess-Zumino term, implying the
eq.(\ref{integ}). This is a highly nontrivial result for a physical
quantity. We are dealing therefore with an ambiguity which is of an unusual
type as compared to the ordinary field theory ambiguities associated to the
genuine ultraviolet divergences, fixed by a set of renormalization
conditions.

For a better understanding of the relationship between $\sigma $ and parity,
we recall here that condensed matter effective models containing $(2+1)$d
fermions are usually defined by implementing, at the lagrangian level, the
parity breaking properties of the system. In particular, information about
transport properties can be obtained from the associated equations of
motion. 
For instance, it is well known that the field modes of the Dirac equation 
\begin{equation}
(iD\!\!\!/+m)\psi =0,  \label{ceq}
\end{equation}
can be quantized to compute a transverse conductance or, equivalently, a
proportionality factor $\sigma $ between charge and flux\cite{NS,eqmt}. This
follows from the fact that the spectrum of the eq.(\ref{ceq}) in the
presence of an external magnetic flux $\Phi $ displays an asymmetry related
to the presence of zero modes, whose degeneracy is $\Phi /(2\pi )$ \cite
{Aharonov-Casher}. As a consequence, the vacuum expectation value of the
charge operator $Q$ receives contributions only from the zero modes, and is
given by\cite{NS,eqmt} 
\begin{equation}
\langle Q\rangle =\frac 12\frac m{|m|}\Phi /(2\pi )\makebox[.5in]{,}Q=\frac 1%
2\int d^2x[\psi ^{\dagger },\psi ].
\end{equation}
This corresponds to a proportionality factor between charge density and
magnetic field (or transverse conductance) $\sigma =\frac 1{4\pi }\frac m{|m|%
}$. Note that the only parity breaking effects are those already present in
the Dirac lagrangian (cf. eq.(\ref{ceq})).

This result may be interpreted as enforcing the criterion of defining the
model at quantum level by not introducing additional parity breaking
effects. For this criterion be well defined, any compatible regularization
scheme should lead to the same transverse conductance. Observe indeed that
the Pauli-Villars or the higher order regularization with $q=0$ agree with
the zero mode calculation.

To further exploit this idea, in the next section we will compute the
induced Chern-Simons coefficient by following the point-splitting method,
which enjoys the property of not introducing additional parity breaking.
Therefore it will provide a nontrivial check for the determination of $%
\sigma $.


\section{The induced Chern-Simons coefficient and the Point-Splitting
regularization}

\label{S:point-splitting} 

Let us start by considering the lagrangian density, defined on a $(2+1)$d
space-time, 
\begin{equation}
{\cal L}=i\overline{\Psi }(x)\gamma ^\mu \partial _\mu \Psi (x)+\overline{%
\Psi }(x)\gamma ^\mu A_\mu \Psi (x)-m\overline{\Psi }(x)\Psi (x),
\label{lagin}
\end{equation}
where $A_\mu $ is an external gauge field, and the matrices $\gamma _\mu $
are defined in terms of the Pauli matrices, $\gamma ^0=\sigma ^3$, $\gamma
^1=i\sigma ^1$, $\gamma ^2=i\sigma ^2$. With this definition $\{\gamma _\mu
,\gamma _\nu \}=2g_{\mu \nu }$ where $g_{\mu \nu }=\mbox{diag}(1,-1,-1)$.

The essence of the point-splitting regularization is to split the product of
local operators by means of the introduction of a small vector $\epsilon _\mu $. It is clear that this
procedure modifies the ultraviolet (short distance) behavior of the theory.
When choosing a particular space-time direction, Lorentz invariance is
broken. However, this symmetry can be recovered by properly averaging the
products over all possible $\epsilon _\mu $ orientations.

The regularized lagrangian density is defined by \cite{becchi-velo} 
\begin{equation}
{\cal L}_{{\rm reg}}=-2i\{\overline{\overline{\Psi }(x+\epsilon )\frac{%
\gamma \cdot \epsilon }{\epsilon ^2}T(e^{i\int_{-1}^1dt\epsilon _\mu A^\mu
(x+\epsilon t)})\Psi (x-\epsilon )}\}-m\overline{\Psi }(x)\Psi (x),
\label{lagsplit}
\end{equation}
where $T$ denotes a time-ordered product with respect to the variable $t$ .
The bar over the first term in eq.(\ref{lagsplit}) represents the average
over the orientations of the vector $\epsilon _\mu $. It is easy to show
that 
\begin{equation}
\lim_{\epsilon \to 0}{\cal L}_{{\rm reg}}={\cal L}.
\end{equation}
We also see that expression (\ref{lagsplit}) preserves gauge and Lorentz
invariance, while a parity transformation only acts by changing the sign of
the fermion mass. In fact, if a parity transformation in the first term of
eq. (\ref{lagsplit}) is considered, one of the components of $\epsilon _\mu $
will change sign, say $\epsilon _1\to -\epsilon _1$. However, this change
has no effect because of the $\epsilon _\mu $ averaging process. Thus, the
regularized lagrangian ${\cal L}_{{\rm reg}}$ shares the same symmetry
properties of the original unregularized lagrangian. 

Following \cite{becchi-velo}, it is convenient to rewrite the action $A=\int
d^3x{\cal L}_{{\rm reg}}(x)$ in momentum space 
\begin{equation}
A=\int \frac{d^3p}{(2\pi )^3}\ \overline{\Psi }(p)S^{-1}(p)\Psi (p)\ +\int 
\frac{d^3p\ d^3q}{(2\pi )^3}\ \overline{\Psi }(p+\frac{\ 1}{\ 2}\ q)\Gamma
(p,q)\Psi (p-\frac{\ 1}{\ 2}\ q),  \label{action}
\end{equation}
where 
\begin{equation}
S^{-1}(p)=-2i\overline{\frac{\gamma \cdot \epsilon }{\epsilon ^2}e^{2ip\cdot
\epsilon }}-m=-\gamma _\mu \frac \partial {\partial p_\mu }(\overline{\frac{%
e^{2ip\cdot \epsilon }}{\epsilon ^2}})-m.  \label{S-1}
\end{equation}
The interactions are encoded in the function 
\begin{equation}
\Gamma (p,q)=\sum_{n=1}^\infty \Gamma ^{(n)}(p,q)  \label{gamma}
\end{equation}
with $\Gamma ^{(n)}$ containing the $n^{{\rm th}}$ power of the gauge field 
\begin{eqnarray}
\Gamma ^{(n)} &=&\frac 1{(2\pi )^{3n}n!}\int (\prod_{j=1}^ndk_j)\delta
(q-\sum_{l=1}^nk_l)T\left\{ \int_{-1/2}^{1/2}\prod_{r=1}^ndt_r\left[
\prod_{s=1}^nA(k_s)\cdot \frac \partial {\partial p}\right]
S^{-1}(p-\sum_{g=1}^nk_gt_g)\right\} .  \nonumber   \\
&& \label{gamma-n}
\end{eqnarray}

We note that a particular average of $[exp(2ip\cdot \epsilon )]/\epsilon ^2$
in eq.\ (\ref{S-1}) corresponds to fixing the form of $S^{-1}(p)$. In the
case where the average enforces Lorentz invariance, we can use the following
ansatz 
\begin{equation}
\frac \partial {\partial p_\nu }(\overline{\frac{e^{2ip\cdot \epsilon }}{%
\epsilon ^2}})=\frac{p^\nu }{f(\mu ,p^2)},  \label{mean value}
\end{equation}
where $\mu $ is a regularization parameter and $f(\mu ,p^2)$ is an arbitrary
analytic function of $p^2$ satisfying 
\begin{equation}
\lim_{\mu \to 0}f(\mu ,p^2)=1.  \label{mu0}
\end{equation}
In terms of $f(\mu ,p^2)$, the regularized propagator reads 
\begin{equation}
S(p)=f(\mu, p^2)\frac{\not{\!p}+m f(\mu, p^2)}{p^2-m^2 f^2(\mu, p^2)} .
\label{propagator}
\end{equation}
Notice that, in the limit where $\mu \to 0$, the free massive Dirac propagator is correctly reobtained. Also, using eqs.(\ref
{gamma-n}) and ( \ref{propagator}) we see that the asymptotic form of $%
\Gamma ^{(n)}$ is 
\begin{equation}
\Gamma _{p\to \infty }^{(n)}(p,q)\approx \frac 1{f(\mu ,p^2)p^{n-1}}.
\label{asympt}
\end{equation}
This equation, together with eq.(\ref{propagator}), implies that all loop
integrals can be made convergent by choosing a function $f(\mu ,p^2)$ with a
fast enough growing behavior as $p\to \infty $. Moreover, from eq.\ (\ref
{asympt}), we can see that among the various $\Gamma ^{(n)}$ ($n=1,\ 2,...$%
), $\Gamma ^{(1)}$ has the lowest decreasing degree as $p\to \infty $.
Therefore, if a particular Feynman diagram $G$ containing $\Gamma ^{(1)}$ as
a vertex corresponds to a finite loop integral, any diagram obtained from $G$
by replacing $\Gamma ^{(1)}\to \Gamma ^{(n)}\ (n=2,\ 3,...)$ will be
associated with a finite loop integral. Thus, in order to determine the
function $f(\mu ,p^2)$ only the diagrams containing the vertex $\Gamma ^{(1)}
$ need to be considered. A simple power counting argument shows that if $%
f(\mu ,p)\approx p^4$ as $p\to \infty $, then all loop integrals turn out to
be finite. These requirements can be fulfilled by choosing, for instance,  $%
f(\mu ,p)=1-(\mu /m)^2p^2+(\mu /m)^4p^4\label{f}$. However, we will proceed
by considering a general form of $f(\mu ,p^2)$ compatible with
the convergence conditions.

As is well known, the Chern-Simons coefficient is completely determined by
the one loop contribution, and is related to the vacuum polarization tensor  $ \Pi_{\mu\nu} $. Let us consider then the vacuum functional 
\begin{equation}
e^{iW(A)}=\langle T[\exp(iA_I)]\rangle_0~~,  \label{vacfunc}
\end{equation}
where $A_I$ is the interaction part of the action. In order to compute the vacuum polarization $ \Pi_{\mu\nu} $, we look for the
quadratic contribution in the gauge fields to the effective action $%
W(A_\mu) $, given by 
\begin{equation}
W^{(2)}=\langle T[(iA_I)^2/2]\rangle_0 ~~.  \label{z2}
\end{equation}
Here, in the expression of $A_I$, all we need to consider is the vertex $%
\Gamma^{(1)}$. The linearized interaction reads 
\begin{equation}
A_I=-2i\int dx\;\overline{\left\{ \overline{\Psi}(x+\epsilon)\frac{\gamma
\cdot \epsilon}{\epsilon^2}\times \left[ i\int_{-1}^{1}dt\; \epsilon\cdot
A(x+\epsilon t)\right] \Psi(x-\epsilon) \right\} }.
\end{equation}
Using this equation and Wick's theorem, we can rewrite $W^{(2)}$ in the
form 
\begin{eqnarray}
W^{(2)}=\frac{\ 1}{\ 2} \int dx_1dx_2\ Tr &&\left\{ \; \overline{(-2i\frac{
\gamma \cdot \epsilon}{\epsilon^2}) \left[ \int_{-1}^{1}dt_1\ \epsilon\cdot
A(x_1+\epsilon t_1)\right]\; \langle\overline{\Psi}(x_1+\epsilon)\Psi(x_2-
\epsilon)\rangle_0\times}\right.  \nonumber \\
&&  \nonumber \\
& &\;\; \left.\overline{\ (-2i\frac{\gamma \cdot \epsilon}{\epsilon^2}
)\left[ \int_{-1}^{1}dt_2\ \epsilon\cdot A(x_2+\epsilon t_2)\right]\;
\langle \overline{\Psi}(x_2+\epsilon)\Psi(x_1-\epsilon)\rangle_0} \right\},
\nonumber \\
&&
\label{z2x1x2}
\end{eqnarray}
where $\langle\overline{\Psi}(x_1+\epsilon)\Psi(x_2-\epsilon)\rangle_0$ and $%
\langle\overline{\Psi}(x_2+\epsilon)\Psi(x_1-\epsilon)\rangle_0$ denote the  free fermion propagators written in coordinate space.

An explicit evaluation of the mean value leads to the following momentum
space result 
\begin{equation}
W^{(2)}=\frac{\ 1}{\ 2}\int \frac{d^3q}{(2\pi)^3}\
A^\mu(q)\Pi_{\mu\nu}(q)A^\nu(-q),  \label{z2pimunu}
\end{equation}
where the regularized vacuum polarization tensor reads 
\begin{eqnarray}
\Pi_{\mu\nu}(q)&=&-Tr\left\{\int \frac{d^3p}{(2\pi)^3}
\int_{-1/2}^{1/2}dt_1dt_2\;\left[ \frac{\partial}{\partial p^\mu}\left(\frac{%
\not{\!u}}{f(\mu,u^2)} \right) \frac{\not{\!p}+mf(\mu,p^2)}{
p^2-m^2f^2(\mu,p^2)}f(\mu,p^2) \right] \times \right.  \nonumber \\
\;&&\;  \nonumber \\
& & \;\;\;\;\;\;\;\;\;\;\;\;\;\;\;\; \left. \left[ \frac{\partial}{\partial
p^\nu}\left(\frac{\not{\!v}}{f(\mu,v^2)}\right) \frac{\not{\!p}\ +\not%
{\!q}+mf(\mu,(p-q)^2)}{(p-q)^2-m^2f^2(\mu,(p-q)^2)} f(\mu,(p-q)^2)\right]
\right\}  \label{pimunu}
\end{eqnarray}
and $u$, $v$ are the vectors 
\begin{eqnarray}
u_\nu&\equiv&p_\nu-q_\nu t_1 \\
v_\nu&\equiv&p_\nu-q_\nu+q_\nu t_2.
\end{eqnarray}
The expression (\ref{pimunu}) can be split into a symmetric and
an antisymmetric part. The symmetric part follows from the parity conserving
terms in the effective action and turns out to be regularization independent; a closed
expression can be found in \onlinecite{jackiw}. On the
other hand, the antisymmetric part is related to the parity breaking terms.
In general, it depends on the regularization through the induced local
Chern-Simons coefficient  
\begin{equation}
\sigma_{(\mu)}=\frac{1}{2}\lim_{q^2\to 0} \epsilon_{\mu\nu\rho} \frac{q^\rho 
}{q^2}\Pi^{\mu\nu}(q^2).  \label{CScoeficient}
\end{equation}
Performing the trace over the spin degrees of freedom, evaluating the integrals over $t$, and taking the limit for small momenta, the regularized induced Chern-Simons coefficient is found to be  
\begin{equation}
\sigma_{(\mu)}=m F(m,\mu),  \label{Pimunuf1}
\end{equation}
where 
\begin{equation}
F(m, \mu)=\frac{2}{3} \int\frac{d^3p}{(2\pi)^3}\frac{4\mu
(p^2/m^2)(1-2\mu^3p^2/m^2)+3}{f(\mu,p^2)[p^2-m^2f^2(\mu,p^2)]^2}.
\label{integral}
\end{equation}
Observe that the integral in the eq.\ (\ref{integral}) is finite for all  values of $\mu$.
Therefore, using eq.\ (\ref{mu0}), we can take the limit $\mu\to 0$ in the
integrand, and perform the momentum integral to obtain 
\begin{equation}
\lim_{\mu \to 0}\sigma_{(\mu)} = \sigma=\frac{1}{4\pi}\frac{m}{|m|}.
\label{sigma}
\end{equation}
We remark that this result does not depend on the explicit
form used for the function $f(\mu,p^2)$. The whole calculation relies on two
properties, namely, at large momenta $f(\mu,p^2)$ grows fast enough so as to
regularize the theory and $\lim_{\mu\to 0}f(\mu,p^2)=1$.

Another important point to be mentioned is that we have taken the limit $\mu
\to 0$ in the integrand of eq.\ (\ref{integral}). This is completely
justified since, as observed before, the integral (\ref{integral}) is 
{\em finite}, for any value of $\mu $. This is not the case, for instance,
of the chiral anomaly in $(3+1)$d where the loop integrals are divergent and it is not possible to exchange the order of the integration with the 
limit $\mu \to 0$\cite{becchi-velo}.


\section{Conclusions}

\label{S:conclusions} 

Following the point-splitting regularization method, we have computed the
induced Chern-Simons coefficient. The latter is seen to be independent of
the particular splitting averaging process, taking the unambiguous value $%
\sigma=\frac{1}{4\pi}\frac{m}{|m|}$.

Our main motivation for the preceding calculation was that of considering a
well defined regularization scheme which does not break parity. In
particular, the point-splitting can be implemented at the lagrangian level 
maintaining, at any stage, translation, Lorentz and {\it small} gauge
symmetry.

This calculation supports the idea that a possible physical criterion for
the determination of the fermionic system can be that of not introducing
additional parity breaking effects. This physical determination is natural
in most condensed matter models that include $(2+1)$d relativistic fermions,
as those describing quantum critical transitions and nodal liquids.

We also note that the same result for $\sigma$ is obtained by counting the
zero modes of the Dirac equation in the presence of an external magnetic
flux \cite{eqmt}.

To some extent, this determination can be compared with similar behavior in $%
(1+1)$d systems. There, because of charge conservation, the natural choice
is not breaking gauge symmetry in most models containing chiral anomalous $%
(1+1)$d fermions, as those describing Luttinger liquids in quantum wires. We
note that the anomalous terms in the effective action for $(2+1)$d fermions
can also be understood \`{a} la Fujikawa \cite{cf00}.

Finally, other physical determinations of the fermionic system cannot of
course be disregarded. Among the alternative possibilities, the requirement
of large gauge invariance, when the Euclidean time coordinate is
compactified to live on a circle, is particularly interesting.

This determination could be relevant when discussing the construction of
Green's functions for the vortex excitations present in the bosonized $(2+1)$%
d theories. These excitations could be created by the introduction of
monopole singularities, which lead to a quantization of the Chern-Simons
coefficient, compatible with large gauge invariance\cite{He-Tei}. This
context would be analogous to that presented in ref.\cite{npb}, where
skyrmion Green's functions are defined by the introduction of instanton
singularities. In both cases, the time compactification can be associated to
fixed boundary conditions around the singularities.

\section{Acknowledgments}

We are indebted to Carlo M.\ Becchi for many discussions on the
point-splitting. We thank C.\ A.\ Linhares, Cesar Fosco, Fidel Schaposnik
and Gerardo Rossini for fruitful comments.

The Conselho Nacional de Desenvolvimento Cient\'{\i}fico e Tecnol\'{o}gico
CNPq-Brazil, the Funda{\c {c}}{\~{a}}o de Amparo {\`{a}} Pesquisa do Estado
do Rio de Janeiro (Faperj), and the SR2-UERJ are acknowledged for the
financial support.

D.\ G.\ B. was partially supported by the National Science Foundation grant
number DMR98-17941, the University of the State of Rio de Janeiro, Brazil
and by the Brazilian agency CNPq through a postdoctoral fellowship.


\end{document}